\documentclass[iop]{emulateapj}
\usepackage[pass,letterpaper]{geometry}
\usepackage{hyperref}

\journalinfo{Accepted to the Astrophysical Journal}
\submitted{Received 2014 December 23; accepted 2015 May 4; published 2015 August 14}

\shorttitle{Chemistry of the $z \gtrsim 10$ Universe}
\shortauthors{Casey \& Schlaufman}

\begin{document}

\title{CHEMISTRY OF THE MOST METAL-POOR STARS IN THE BULGE AND THE
$z\gtrsim10$ UNIVERSE\footnote{This paper includes data gathered with the
6.5 m Magellan Telescopes located at Las Campanas Observatory, Chile.}}

\author{Andrew R.\ Casey\altaffilmark{1} and
Kevin C.\ Schlaufman\altaffilmark{2,3}}
\affil{
$^1$ Institute of Astronomy, University of Cambridge,
Madingley Road, Cambridge CB3 0HA, UK; arc@ast.cam.ac.uk\\
$^2$ Kavli Institute for Astrophysics and Space Research, Massachusetts
Institute of Technology, Cambridge, MA 02139, USA; kschlauf@mit.edu}
\altaffiltext{3}{Kavli Fellow.}

\begin{abstract}

Metal-poor stars in the Milky Way are local relics of the epoch of the
first stars and the first galaxies.  However, a low metallicity does
not prove that a star formed in this ancient era, as metal-poor stars
form over a range of redshift in different environments.  Theoretical
models of Milky Way formation have shown that at constant metallicity,
the oldest stars are those closest to the center of the Galaxy on the
most tightly-bound orbits.  For that reason, the most metal-poor stars in
the bulge of the Milky Way provide excellent tracers of the chemistry of
the high-redshift universe.  We report the dynamics and detailed chemical
abundances of three stars in the bulge with $\mathrm{[Fe/H]}\lesssim-2.7$,
two of which are the most metal-poor stars in the bulge in the literature.
We find that with the exception of scandium, all three stars follow
the abundance trends identified previously for metal-poor halo stars.
These three stars have the lowest [\ion{Sc}{2}/Fe] abundances yet seen in
$\alpha$-enhanced giant stars in the Galaxy.  Moreover, all three stars
are outliers in the otherwise tight [\ion{Sc}{2}/Fe]--[\ion{Ti}{2}/Fe]
relation observed among metal-poor halo stars.  Theoretical models
predict that there is a 30\% chance that at least one of these stars
formed at $z\gtrsim15$, while there is a 70\% chance that at least one
formed at $10\lesssim\!z\!\lesssim15$.  These observations imply that
by $z\sim10$, the progenitor galaxies of the Milky Way had both reached
$\mathrm{[Fe/H]}\sim-3.0$ and established the abundance pattern observed
in extremely metal-poor stars.

\end{abstract}

\keywords{Galaxy: bulge --- Galaxy: halo --- Galaxy: stellar content ---
          stars: abundances ---  stars: kinematics and dynamics  ---
          stars: Population II}

\section{Introduction}

The first stars are thought to form at $z \gtrsim 15$, with the first
galaxies following at $z \sim 10$ \citep[e.g.,][]{bro99,abe02,bro11}.
The chemical abundances of these first galaxies are unknown.  If those
abundances could be measured, then they would constrain the properties
of metal-free Population III stars, the early chemical evolution of
galaxies, and the reionization of the universe.  Metal-poor stars in the
Milky Way provide a local link to this high-redshift universe through
the elemental abundances of their photospheres.

As the number of known metal-poor stars with detailed chemical
abundance measurements has grown, it has become possible to
homogeneously analyze large samples to search for subtle trends
\citep[e.g.,][]{cay04,bon09,nor13a,nor13b,yon13a,yon13b,roe14}.  It is
tempting to assert that these metal-poor stars in the halo are the direct
descendants of the first stars.  This is not necessarily the case though,
as metal-poor stars form over a range of redshift in halos of varying
mass and environment.  Likewise, stars at a given redshift form with
a range of metallicity.  The examination of other properties beyond
metallicity are therefore necessary to identify the stars in the Milky
Way that formed at the highest redshifts.

\citet{tum10} showed that because galaxies form from the inside-out, the
oldest stars at a given metallicity are found near the center of a halo
on the most tightly-bound orbits.  Indeed, near the center of a Milky
Way-analog a large fraction of stars with $-3 \lesssim \mathrm{[Fe/H]}
\lesssim -2$ formed at $z \gtrsim 6$, while 20--40\% of stars with
$-4 \lesssim \mathrm{[Fe/H]} \lesssim -3$ formed at $10 \lesssim z
\lesssim 15$.  Consequently, the metal-poor stellar population in the
inner few kpc of the Galaxy---the bulge---is the best place to search
for truly ancient stars, including low-mass Population III stars that
may have survived to the present day.

Large-scale spectroscopic surveys of the bulge have shown that
while metal-poor stars in the bulge are quite rare, they do exist.
The Abundances and Radial Velocity Galactic Origins (ARGOS) survey of
\citet{fre13} and \citet{nes13} identified 16 stars with $\mathrm{[Fe/H]}
\lesssim -2.0$ in a sample of 14,150 stars within 3.5 kpc of the Galactic
center.  The most metal-poor star in their sample has $\mathrm{[Fe/H]}
\approx -2.6$.  As part of the third phase of the Sloan Digital Sky
Survey, the Apache Point Observatory Galactic Evolution Experiment
(APOGEE) collected $H$-band spectra for 2,403 giants stars in outer
bulge fields and identified two stars with $\mathrm{[Fe/H]} \approx -2.1$
\citep{gar13}.

Ground-based objective prism surveys for metal-poor stars in the
bulge are impractical due to crowding and strong absolute and
differential reddening.  For this reason, searches for metal-poor
stars have historically avoided the inner regions of our own Galaxy.
Recently though, the Extremely Metal-poor BuLge stars with AAOmega
(EMBLA) survey has successfully used narrow-band SkyMapper $v$-band
photometry \citep{bes11} in the \ion{Ca}{2} H \& K region to pre-select
candidate metal-poor stars for follow-up spectroscopy.  In a sample
of more than 8,600 stars, \citet{how14} found in excess of 300 stars
with $\mathrm{[Fe/H]} \lesssim -2.0$---including four stars with $-2.7
\lesssim \mathrm{[Fe/H]} \lesssim -2.5$.  Still, strong absolute and
significant differential reddening limits the efficiency of near UV
based selections for metal-poor stars in the bulge and restricts their
applicability to outer-bulge regions.

In \citet{sch14}, we described a new technique to identify candidate
metal-poor stars using only near-infrared 2MASS and mid-infrared
\textit{WISE} photometry \citep{skr06,wri10,mai11}.  Our infrared
selection is well suited to a search for metal-poor stars in the bulge,
as it is minimally affected by crowding or reddening.  We found that
more than 20\% of the candidates selected with our infrared selection are
genuine very metal-poor (VMP) stars with $-3.0 \lesssim \mathrm{[Fe/H]}
\lesssim -2.0$.  Another 2\% of our candidates are genuine extremely
metal-poor (EMP) stars with $-4.0 \lesssim \mathrm{[Fe/H]} \lesssim -3.0$.
In a sample of 90 metal-poor candidates---selected with only an apparent
magnitude cut to be high in the sky from Las Campanas in the first half of
the year---we identified three stars with $-3.1 \lesssim \mathrm{[Fe/H]}
\lesssim -2.7$ within 4 kpc of the Galactic center.  Two of these stars
are the most metal-poor stars in bulge in the literature, while the
third is comparable to the most metal-poor star from \citet{how14}.

Because these stars are both tightly bound to the Galaxy and very
metal-poor, they are likely to be among the most ancient stars identified
to this point.  For that reason, their detailed abundances provide
clues to the chemistry of the first galaxies in the $z \gtrsim 10$
universe, beyond those already identified in more metal-poor halo
stars.  These stars all have apparent magnitudes $V \lesssim 13$,
making them unusually bright for stars at the distance of the bulge.
Their bright apparent magnitudes enable a very telescope-time efficient
exploration of the $z \gtrsim 10$ Universe.  We describe the collection
of the data we will subsequently analyze in Section \ref{sec:data}.
We detail the determination of distances and orbital properties, stellar
parameters, and chemical abundances of these three stars in Section
\ref{sec:analysis}.  We discuss our results and their implications in
Section \ref{sec:discussion}, and we summarize our findings in Section
\ref{sec:conclusions}.

\section{Data Collection}\label{sec:data}

We initially selected these stars as candidates according to
criteria (1)--(4) from Section 2 of \citet{sch14}: $0.45 \leq J-H
\leq 0.6$, $W3 > 8$, $-0.04 \leq W1-W2 \leq 0.04$, and $J-W2 > 0.5$.
We give astrometry and photometry for each star in Table~\ref{tbl-1}.
We confirmed their metal-poor nature using low-resolution spectroscopy
from Gemini South/GMOS-S \citep{hoo04}\footnote{Programs GS-2014A-A-8
and GS-2014A-Q-74.} in service mode during March and April of 2014.
Our Gemini South/GMOS-S follow-up spectroscopy was not focused on
candidates in the bulge, so the discovery of these stars in the bulge was
not predetermined by our survey strategy.  We used the Magellan Inamori
Kyocera Echelle (MIKE) spectrograph \citep{ber03} on the Clay Telescope at
Las Campanas Observatory on 2014 June 21--22 to obtain high-resolution,
high signal-to-noise (S/N) spectra suitable for a detailed chemical
abundance analysis.  We observed all three stars in 0\farcs5~seeing at
airmass $<\!\!1.01$ with exposure times in the range 390--590 seconds.
The total exposure time for all three sources combined was less than
24 minutes.  Including overheads, our Magellan/MIKE observations for all
three stars were completed in about 30 minutes.  We used the 0\farcs7~slit
and the standard blue and red grating azimuths, yielding spectra between
332 nm and 915 nm with resolution $R \approx 41,\!000$ in the blue and $R
\approx 35,\!000$ in the red. The resultant spectra have S/N $\gtrsim 50$
pixel$^{-1}$ at 400 nm and S/N $\gtrsim 100$ pixel$^{-1}$ at 600 nm.

To obtain proper motions for each star, we cross-matched
with both the UCAC4 and SPM4 proper motion catalogs using
\texttt{TOPCAT}\footnote{\url{http://www.star.bris.ac.uk/~mbt/topcat/}}\citep{zac13,gir11,tay05}.
We list both sets of proper motions for our sample in Table~\ref{tbl-2}.

\section{Analysis}\label{sec:analysis}

We reduced the spectra using the
\texttt{CarPy}\footnote{\url{http://code.obs.carnegiescience.edu/mike}}
software package \citep{kel03,kel14}.  We continuum-normalized individual
echelle orders using spline functions before joining them to form a
single contiguous spectrum.  We estimate line-of-sight radial velocities
by cross-correlating each spectrum with a normalized rest-frame spectrum
of the well-studied metal-poor giant star {HD 122563}. We use the measured
radial velocities to place the spectra in the rest-frame of the star.

\subsection{Distances \& Dynamics}

To determine the distances between the sun and each star in our sample,
we use the scaling relation
\begin{eqnarray}
L/L_{\odot} & = & (R/R_{\odot})^2 (T_{\mathrm{eff}}/T_{\mathrm{eff,\odot}})^4, \\
            & = & (M/M_{\odot}) (g/g_{\odot})^{-1} (T_{\mathrm{eff}}/T_{\mathrm{eff,\odot}})^4.
\end{eqnarray}
Taking their characteristic mass as $0.8~M_{\odot}$, the bolometric
luminosity $L$ of our stars can be approximated as
\begin{eqnarray}
\log{\left(L/L_{\odot}\right)} & = & \log{0.8} - (\log{g} - 4.44) + 4\log{\left(T_{\mathrm{eff}}/5777~\mathrm{K}\right)}.\nonumber\\\label{eq-lum}
\end{eqnarray}
We then use Equation \ref{eq-lum} and the stellar parameters from
\citet{sch14} listed in Table~\ref{tbl-3} to determine $L$.  We use
a 10 Gyr, ${\mathrm{[Fe/H]} = -2.5}$, and ${[\alpha/{\rm Fe}] =
+0.4}$ Dartmouth isochrone to convert $L$ into an absolute $W1$-band
magnitude $M_{W1}$ \citep{dot08}.  Given the  available photometry,
$W1$ is least affected by extinction.  We de-redden the observed $W1$
magnitudes using the \cite{sch98} dust maps as updated in \citet{sch11}
along with the \citet{ind05} infrared extinction law.  The distance
modulus $W1-M_{W1}$ then yields $d_{\odot}$, the approximate distance
of each star from the sun.  Assuming the distance to the Galactic center
is $R_{0} = 8.2 \pm 0.4$ kpc \citep[e.g.,][]{bov09}, we can then compute
$d_{\mathrm{gc}}$, the approximate distance of each star from the Galactic
center.  We perform a Monte Carlo simulation to account for the random
observational uncertainties in $W1$, $A_{W1}$, $T_{\mathrm{eff}}$,
$\log{g}$, and $R_{0}$.  We sample 10,000 realizations from the
uncertainty distributions for each quantity and compute $d_{\odot}$
and $d_{\mathrm{gc}}$ for each realization. We give both distance
estimates and their random uncertainties in the first two columns of
Table~\ref{tbl-4}.  All three stars have $d_{\mathrm{gc}} \lesssim 4$ kpc.

We compute the Galactic orbits of each star in our sample using the
\texttt{galpy} code\footnote{\url{http://github.com/jobovy/galpy} and
described in \citet{bov15}}, with initial conditions set by the observed
heliocentric radial velocities and proper motions in Table~\ref{tbl-2}
and estimated $d_{\odot}$ values from Table~\ref{tbl-4}.  Following
\citet{bov12}, we model the Milky Way's potential as the superposition
of a Miyamoto-Nagai disk with a radial scale length of 4 kpc and a
vertical scale height of 300 pc, a Hernquist bulge with a scale radius
of 600 pc, and a Navarro-Frenk-White halo with a scale length of 36 kpc
\citep{miy75,her90,nav96}.  We assume that the Miyamoto-Nagai disk,
the Hernquist bulge, and the Navarro-Frenk-White halo respectively
contribute 60\%, 5\%, and 35\% of the rotational support at the solar
circle.  We integrate the orbits for 200 orbital periods and derive
the pericenters $r_{\mathrm{peri}}$, apocenters $r_{\mathrm{ap}}$,
and eccentricities $e$.  We perform a Monte Carlo simulation to
account for the random observational uncertainties in $d_{\odot}$,
$v_{\mathrm{hel}}$, $\mu_{\alpha} \cos{\delta}$, and $\mu_{\delta}$.
We sample 1,000 realizations from the uncertainty distributions for
each quantity and use those data as input to an orbital integration.
In an attempt to quantify the systematic uncertainties that result from
the input proper motion measurements, we include in Table~\ref{tbl-4}
orbital properties and uncertainties estimated using both UCAC4 and SPM4
proper motions.

\subsection{Stellar Parameters}

We estimate stellar parameters by classical excitation and ionization
balance using unblended \ion{Fe}{1} and \ion{Fe}{2} lines.  Following
the process described in \citet{cas14}, we measure equivalent widths
of individual absorption lines from the rest-frame spectra by fitting
Gaussian profiles.  We visually inspect all lines for quality, and discard
blended or low-significance measurements. For these analyses, we assume
transitions are in local thermodynamic equilibrium (LTE) and employ the
plane-parallel 1D $\alpha$-enhanced model atmospheres from \citet{cas04}.
We use the atomic data compiled by \citet{roe10}\footnote{We used the
correct transition probabilities for \ion{Sc}{2} from \citet{law89}
that were misstated in \citet{roe10}.}, the \citet{asp09} solar chemical
composition, and the February 2013 version of MOOG to calculate line
abundances and synthesize spectra \citep{sne73,sob11}.  We require four
conditions to be simultaneously met for a converged set of stellar
parameters: zero trend in \ion{Fe}{1} line abundance with excitation
potential, zero trend in \ion{Fe}{1} line abundances with reduced
equivalent width, equal mean \ion{Fe}{1} and \ion{Fe}{2} abundances,
and that the mean [\ion{Fe}{1}/H] abundance must match the input model
atmosphere abundance [M/H].  In practice we accepted solutions where the
slopes had magnitudes less than $10^{-3}$ and the absolute abundance
differences were less than $10^{-2}$ dex.  Our estimated stellar
parameters are provided in Table~\ref{tbl-3}.

\begin{figure}
%\epsscale{1.0} % aastex
\plotone{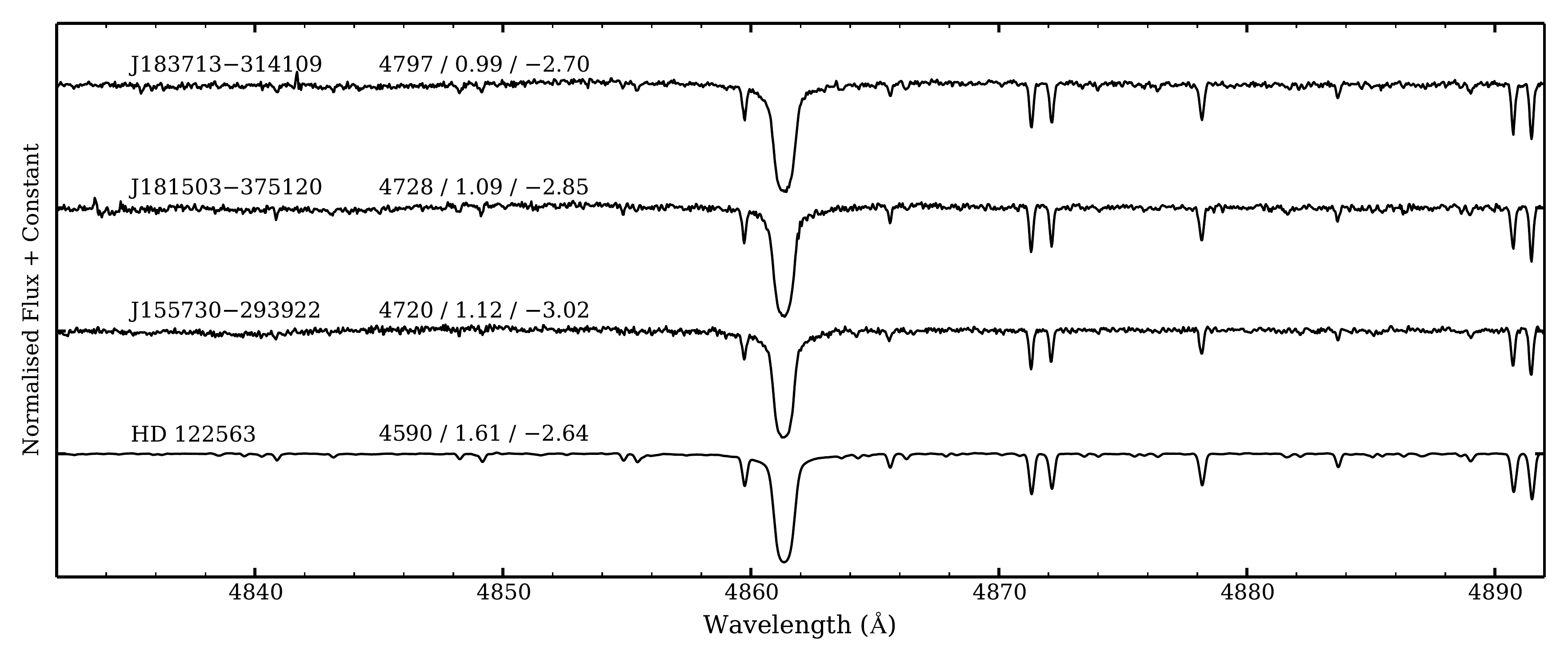}
\caption{Continuum-normalized Magellan/MIKE spectra for the three
metal-poor stars in the bulge along with the well-studied metal-poor
giant {HD\,122563}. The spectra are centered around the H-$\beta$ line,
highlighting the similarity between {HD\,122563} and our metal-poor stars
in the bulge. We indicate the stellar parameters $T_{\mathrm{eff}}$,
$\log{g}$, and $\mathrm{[Fe/H]}$ for each star, with the parameters for
{HD\,122563} from \citet{jof14}.\label{fig01}}
\end{figure}

To verify our spectroscopically-derived effective temperatures,
we calculate effective temperatures using color--temperature
relations for 2MASS $J-K_{s}$ and APASS/2MASS $V-K_{s}$ colors.
We use the \citet{sch98} dust maps as updated by \citet{sch11} to
account for reddening in both colors.  However, we find that our
photometric temperatures are as much as $600$\,K hotter than our
spectroscopically-derived quantities.  To explore the reason for this
discrepancy, we also estimate effective temperatures by comparing
the observed Balmer lines with synthetic spectra from \citet{bar03}.
Our analysis of the H-$\beta$ profile suggests effective temperatures
between $4600$\,K and 4800\,K for all three stars, in excellent
agreement with our excitation-ionization balance measurements.  As we
show qualitatively in Figure~\ref{fig01}, our observed spectra are very
similar to the well-studied metal-poor giant star {HD\,122563}.  Given our
independent effective temperature estimates, and since {HD\,122563}
is a red giant branch star with $T_{\mathrm{eff}} = 4590$\,K, $\log{g}
= 1.61$, and $\mathrm{[M/H]} = -2.64$ \citep{jof14}, we are confident
in our derived spectroscopic effective temperatures.  Moreover, we
observe repeated saturated interstellar \ion{Na}{1} D absorption lines
in our data.  These lines are indicative of multiple optically-thick
gas clouds along the line-of-sight, each with distinct velocities.
For these reasons, we assert that the discrepancy between photometric
and spectroscopic temperatures is likely due to poorly-characterized
reddening in the outer bulge region.  Given the spectral resolution
and S/N ratios of our data, we estimate that the uncertainties in our
spectroscopically-derived stellar parameters are about $100$\,K in $T_{\rm
eff}$, 0.2\,dex in $\log{g}$, 0.1\,dex in [Fe/H], and 0.1\,km s$^{-1}$
in microturbulence ($\xi$).

\begin{figure}
%\epsscale{1.0} % aastex
\plotone{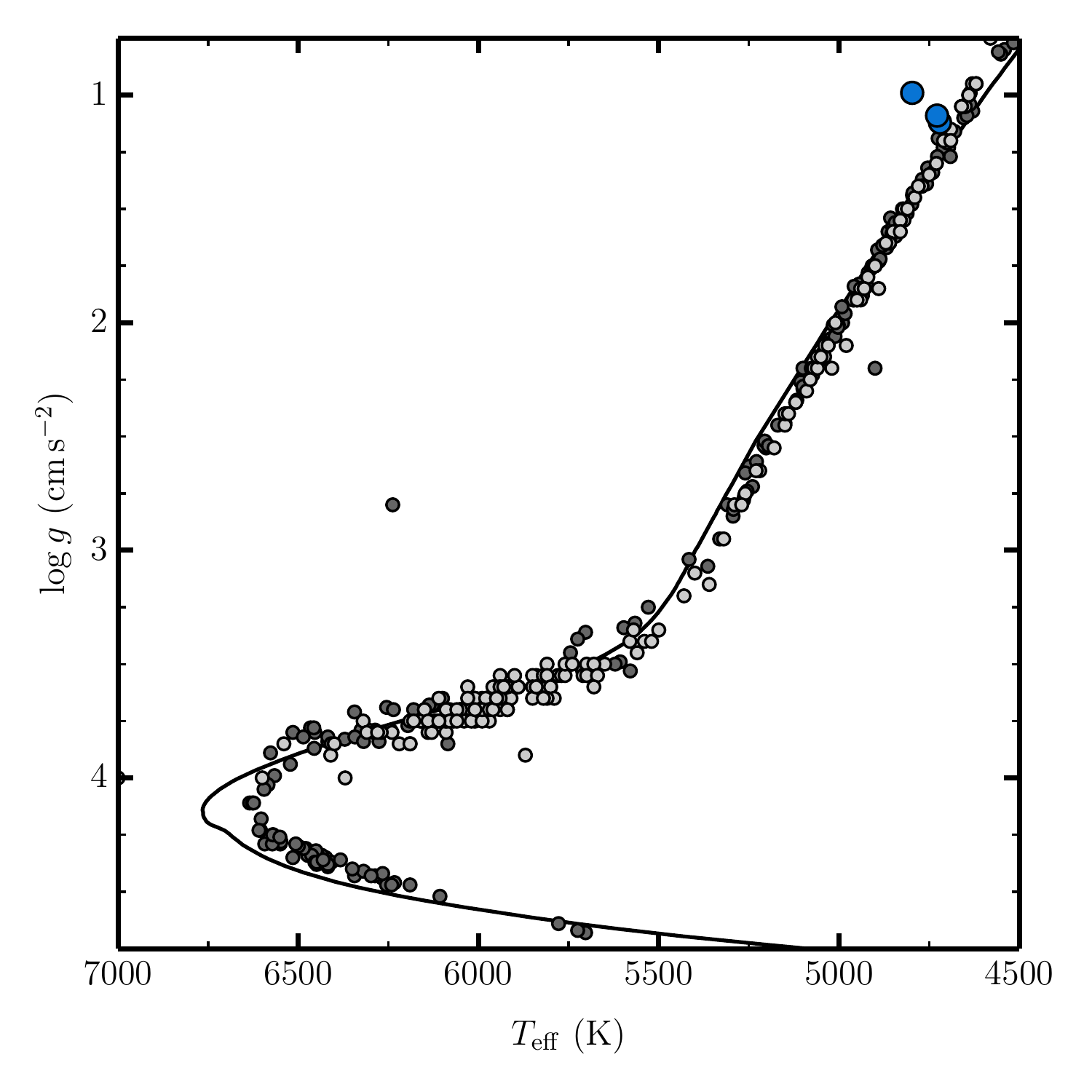}
\caption{Measured effective temperatures and surface gravities of
metal-poor stars.  We plot the locations of our three bulge metal-poor
stars in blue.  For comparison, we plot halo stars with $\mathrm{[Fe/H]}
\lesssim -2.0$ from \citet{yon13a} in dark gray and from \citet{roe14}
in light gray.  We only plot stars from \citet{roe14} where the surface
gravity was derived from isochrones.  The solid line is a 12 Gyr,
$\mathrm{[Fe/H]} = -2.5$, and [$\alpha$/Fe] $= +0.4$ Dartmouth isochrone.
The \citet{yon13a} sample deviates from the displayed isochrone at the
main sequence turn-off because \citet{yon13a} used the Y$^2$ isochrones
\citep{dem04} to determine stellar parameters. Even though we measured
our stellar parameters from excitation and ionization balance, our stars
are largely in agreement with isochrone-derived surface gravities in
the comparison samples.\label{fig02}}
\end{figure}

We note that our stellar parameters ($T_{\rm eff}$, $\log{g}$, [Fe/H],
$\xi$) would change if we used different model atmospheres or included
a proper treatment of non-LTE effects.  For metal-poor giants, the
non-LTE treatment would increase the mean \ion{Fe}{1} line abundance
by about 0.1\,dex and result in higher surface gravities for a given
effective temperature.  As an example, \citet{jof14} reports a slightly
cooler temperature and higher surface gravity for HD 122563 than we find
for our three stars.  However, in that study $T_{\rm eff}$ and $\log{g}$
were not derived by excitation and ionization equilibrium.  Instead, they
were fixed by bolometric temperature and angular diameter measurements
from \citet{cre12}.  With the stellar parameters fixed, \citet{jof14}
noted that HD 122563 showed the largest abundance imbalance of \ion{Fe}{1}
and \ion{Fe}{2} lines in their sample.  This indicates the the application
of the equilibrium method in LTE tends towards a different set of stellar
parameters.  In Figure~\ref{fig02} we plot our stars alongside giant star
(i.e., $\log{g} \lesssim 3.0$) comparison samples from \citet{yon13a}
and \citet{roe14}.  Although these authors estimated surface gravities
directly from isochrones, our stellar parameters are comparable to
their determinations.  Consequently, we are confident of our stellar
parameter estimates.

\subsection{Detailed Abundances}

Our high-resolution, high S/N Magellan/MIKE spectra allow us to
measure the abundances of many light, odd-Z, $\alpha$, Fe-peak, and
neutron-capture elements.  For most elements, we determine individual
line abundances from the measured equivalent widths of clean, unblended
atomic lines.  We take a synthesis approach for molecular features (e.g.,
CH), doublets (e.g., Li), or atomic transitions with significant hyperfine
structure and/or isotopic splitting (namely Sc, V, Mn, Co, Cu, Ba, La,
and Eu).  We use molecular data (CH) from \citet{mas14}.  Our hyperfine
structure and isotopic splitting data come from \citet{kur95} for Sc, V,
Mn, Co, and Cu, from \citet{bie99} for Ba, and from \citet{law01a,law01b}
for La and Eu.  We assume standard solar system isotopic fractions as
collated by \citet{and89}.  We report our equivalent width measurements
in Table~\ref{tbl-5} and our derived abundances in Table~\ref{tbl-6}.

We estimate lithium abundances through synthesis of the Li doublet at
$\lambda$6707. This feature is quite weak in our spectra.  However, the
abundances we obtain are typical for stars at the tip of the red giant
branch.  We synthesize the $G$-band molecular feature at $\lambda$4323 to
estimate carbon abundances.  None of our stars are carbon enhanced
by the \citet{bee05} definition of [C/Fe] $\gtrsim +1.0$.  On the
other hand, one of our stars is carbon enhanced by the \citet{aok07}
definition that takes stellar evolutionary effects into account. In
either case, there is not much carbon present in the photospheres of
our stars---[C/Fe] ranges from $-0.61$ in J183713-314109 to $+0.15$ in
J181503-375120.  We measure potassium abundances from equivalent widths
of the strong \ion{K}{1} transitions at $\lambda$7664 and $\lambda$7698.
Given the radial velocities of our targets, these \ion{K}{1} lines were
mostly separated from the telluric A-band feature near $\lambda$7600.
We detected \ion{Na}{1} in all three stars and derive abundances from
the strong $\lambda$5889 and $\lambda$5895 transitions.  We measure
\ion{Al}{1} from the $\lambda$3961 feature.

\begin{figure}
\epsscale{0.5} % aastex
\plotone{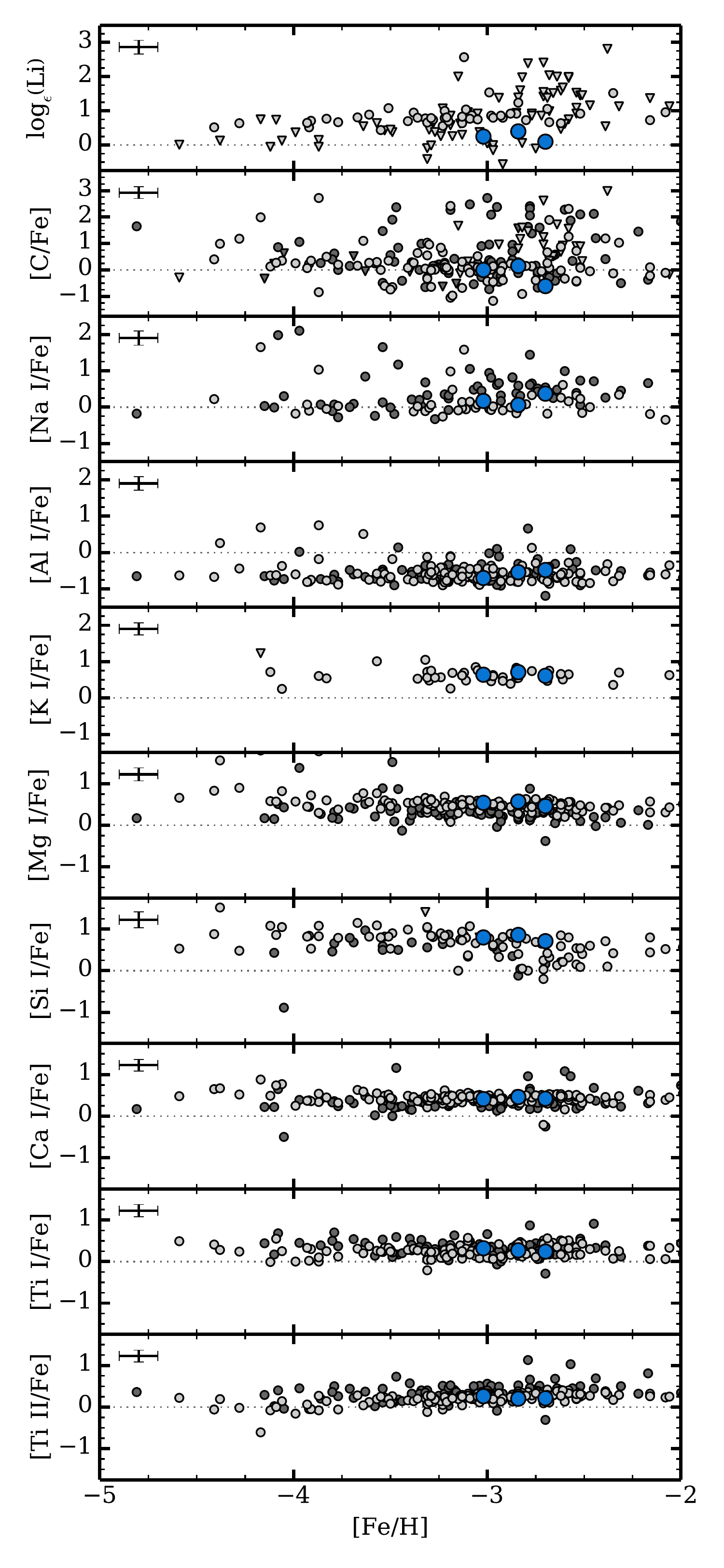} % aastex
\caption{Chemical abundances of Li, C, odd-Z (Na, Al, K), and
$\alpha$-elements (Mg, Si, Ca, Ti) for metal-poor stars.  We plot our
three bulge metal-poor stars in blue, the \citet{yon13a} giant (i.e.,
$\log{g} < 3$) comparison sample in dark gray, and the \citet{roe14} giant
sample in light gray.  Measurements are indicated by circles and upper
limits are shown as triangles.  Typical uncertainties are given.  We plot
here the [\ion{K}{1}/Fe] abundance ratios from \citet{roe14} without
correcting for non-LTE effects, such  that they are comparable with our
analysis.  All other abundances from \citet{yon13a} and \citet{roe14}
shown here also assume LTE.  Although the $y$-axis scale varies in each
panel to accommodate the dynamic range of each abundance, the minor tick
marks are spaced at $0.25$\,dex in all panels.\label{fig03}}
\end{figure}

All three stars appear $\alpha$-enhanced (Mg, Ti, Si, and Ca). On average,
the $\alpha$-element abundances of these three metal-poor stars in the
bulge are similar to those observed in large samples of halo metal-poor
giant stars \citep[e.g.,][]{cay04,yon13a,roe14}.  [Mg/Fe] varies between
$+0.46$ and $+0.57$, while [Ca/Fe] changes marginally from $+0.41$
to $+0.47$.  However, in all stars we find that [\ion{Ti}{1}/Fe] and
[\ion{Ti}{2}/Fe] are slightly lower than the other $\alpha$-elements,
between [Ti/Fe]$ = +0.22$ and $+0.29$ (Figure~\ref{fig03}).  In all
stars, the mean abundances of neutral and ionized Ti transitions agree
within 0.03--0.08\,dex.  We measure [\ion{Si}{1}/Fe] abundances from
the $\lambda$3905 transition, yielding [\ion{Si}{1}/Fe] abundance ratios
between $+0.71$ and $+0.86$.

There are a large number of Fe-peak transitions available in our spectra:
\ion{Sc}{2}, \ion{V}{1}, \ion{Cr}{1} \& \ion{Cr}{2}, \ion{Mn}{1},
\ion{Co}{1}, \ion{Ni}{1}, \ion{Cu}{1}, and \ion{Zn}{1}. While Sc, V, Cr,
Mn, Co, Ni, and Zn are clearly measurable in all stars from multiple
unblended lines, we do not detect \ion{Cu}{1} in J155730-293922 or
J183713-314109. Instead, we provide upper limits for \ion{Cu}{1} from the
$\lambda$5105 transition.  We also report a low-significance detection
of \ion{Cu}{1} in J181503-375120 of {[\ion{Cu}{1}/Fe] $= -0.51$}. Our
Fe-peak abundance ratios generally follow the mean halo abundance
trends observed by other authors in giant stars of similar metallicity
\citep[e.g.,][]{cay04,yon13a,roe14}. We find that [\ion{Si}{1}/Fe],
[\ion{Sc}{2}/Fe], and [\ion{Mn}{1}/Fe] abundances are at the extremes
of the abundance distribution observed in halo metal-poor giant stars.
We show this in Figure~\ref{fig05} and explore possible explanations
for these observations in Section \ref{sec:discussion}.

\begin{figure}
\epsscale{0.7} % aastex
\plotone{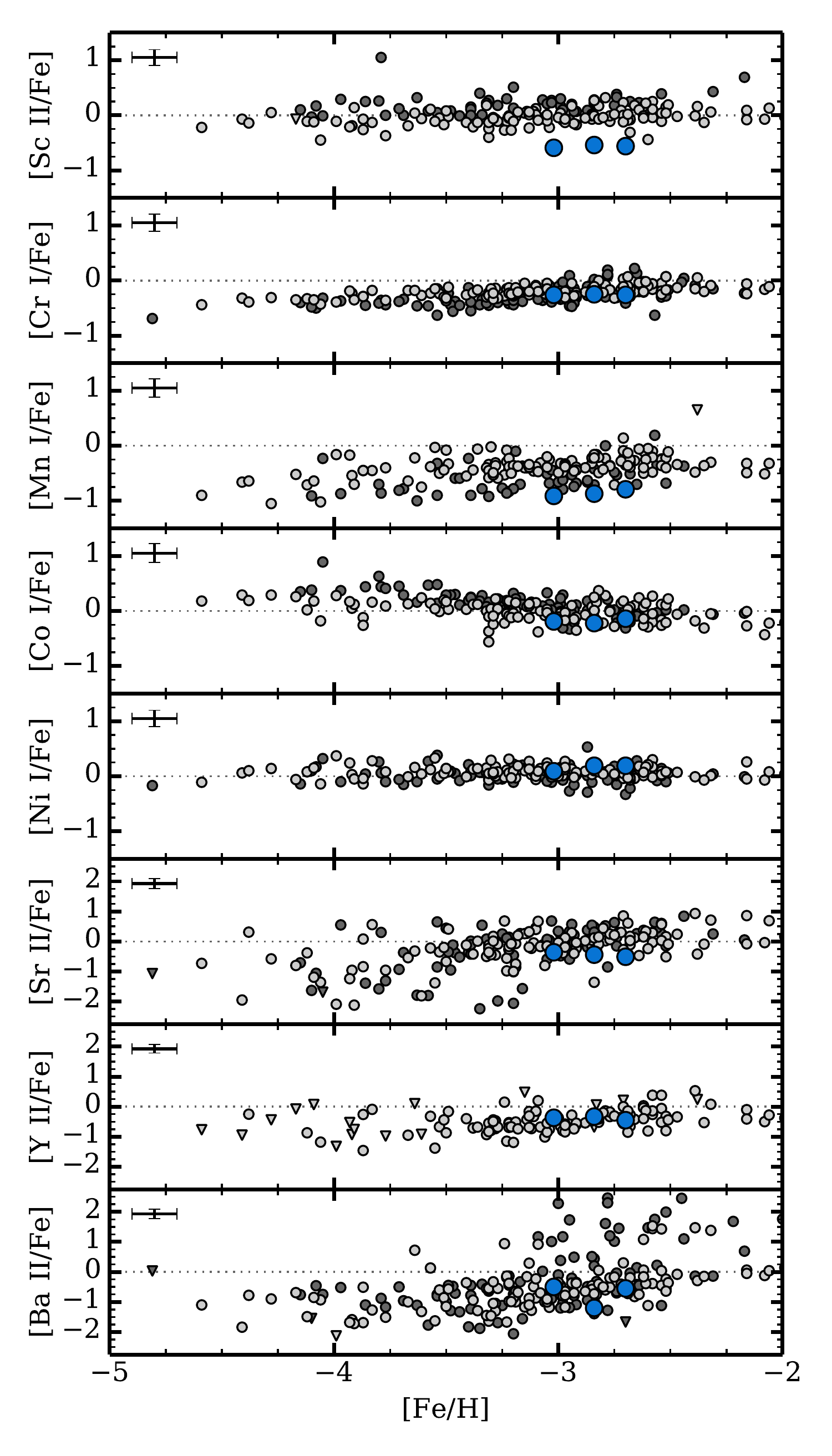}
\caption{Chemical abundances of Fe-peak and neutron-capture elements
for metal-poor giant stars.  We plot our three bulge metal-poor stars
in blue, the \citet{yon13a} comparison sample in dark gray, and the
\citet{roe14} sample in light gray.  Measurements are indicated with
circles and upper limits are shown as triangles. Typical uncertainties
are given.\label{fig04}}
\end{figure}

\begin{figure}
%\epsscale{1.0} % aastex
\plotone{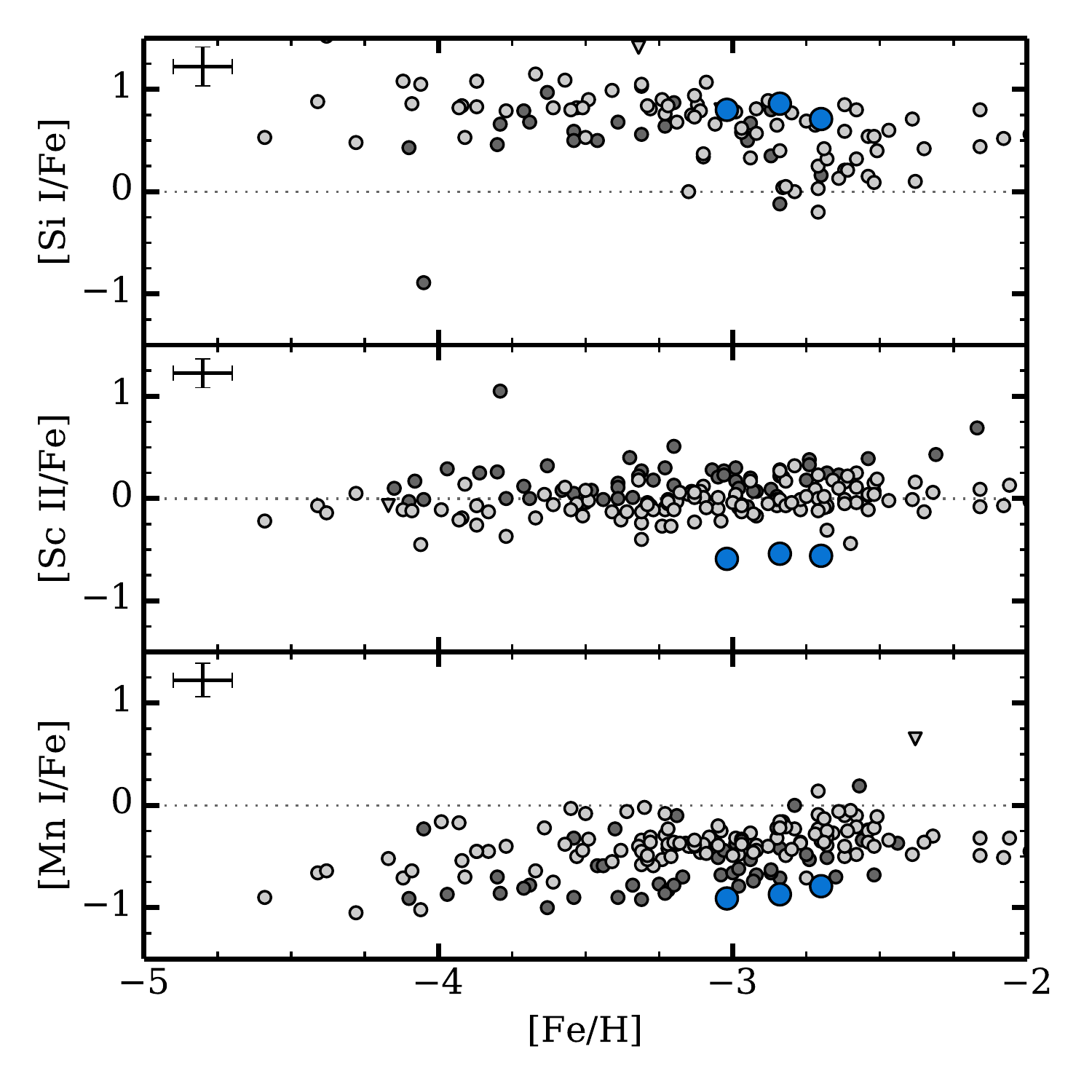}
\caption{Chemical abundances of \ion{Si}{1}, \ion{Sc}{2}, and \ion{Mn}{1}
with respect to Fe for metal-poor giant stars.  We plot our three bulge
metal-poor stars in blue, the \citet{yon13a} comparison sample in dark
gray, and the \citet{roe14} sample in light gray.  Measurements are
indicated with circles and upper limits are shown as triangles. Typical
uncertainties are given.  While our abundances are generally in good
agreement with those measured in halo metal-poor stars, we find that
silicon, scandium, and manganese are all on the extremes of the halo
distribution.\label{fig05}}
\end{figure}

We measure elemental abundances from the first (Sr and Y) and second
(Ba) neutron-capture peaks. We do not detect Eu or La in our targets, and
therefore we report upper limits for these elements in Table~\ref{tbl-6}.
Sr and Y have a common nucleosynthetic pathway, and we observe comparable
abundance ratios for these elements in all three stars.  As we show in
Figure~\ref{fig04}, all of our measured neutron-capture abundances are
indistinguishable from the abundances observed in halo metal-poor giant
stars \citep{yon13a,roe14}.

The uncertainties in chemical abundances are dominated by systematics,
principally due to the uncertainties in determining stellar parameters.
We vary the stellar parameters of each star by the estimated
uncertainties and calculate the resulting change in abundances.  We give
the sign and magnitude of these effects in Table~\ref{tbl-7}, along with
the quadrature sum of systematic uncertainties.  Due to a lack of lines
for some elements, we adopt a minimum random uncertainty of 0.1\,dex.
We estimate total uncertainties as the quadrature sum of random and
systematic uncertainties, which we list in Table~\ref{tbl-7}. For
uncertainties in [X/Fe] abundance ratios (e.g., as shown in
Figures~\ref{fig03}--\ref{fig05}), we adopt the quadrature sum of the
total uncertainties in [X/H] and [\ion{Fe}{1}/H].

\section{Discussion}\label{sec:discussion}

Our initial survey was not targeted at the bulge, so we are observing
all three stars at random orbital phases.  Since a star on a radial
orbit spends most of its orbit near apocenter, there is a strong prior
that we are observing all three stars close to apocenter.  Our estimated
Galactocentric distances and orbital parameters for the three stars listed
in Table~\ref{tbl-4} securely place J155730-293922 and J183713-314109 in
the bulge on tightly bound orbits.  In both cases, the currently-observed
Galactocentric distances are consistent with the idea that both stars
are near apocenter.  At the same time, the differences in proper motion
reported by UCAC4 and SPM4 deviate by up to 3-$\sigma$.  It seems clear
that the quoted random proper motion uncertainties are not representative
of the total uncertainties including the contribution from systematics.
Both stars have $V \lesssim 13$ and have had their proper motions
matched to the correct 2MASS sources, so the discrepancy is not due
to faintness or misidentification.  Nevertheless, the range in proper
motions reported by UCAC4 and SPM4 should be an approximation of the
effect of the unreported systematic uncertainties.  Since both UCAC4 and
SPM4 place J155730-293922 and J183713-314109 on tightly bound orbits,
there is no reason to reject the idea that they are indeed tightly bound.
We therefore argue that since J155730-293922 and J183713-314109 are
metal-poor, located near the center of the Galaxy, and on tightly bound
orbits, they are likely to be truly ancient stars according to the
analysis described in \citet{tum10}.

On the other hand, the orbital parameters listed in Table~\ref{tbl-4}
for the star J181503-375120 suggest that it may be a halo star on a
very eccentric orbit.  Both UCAC4 and SPM4 agree that $\mu_{\alpha}
\cos{\delta} \approx 20$ mas yr$^{-1}$ with high significance,
indicating a substantial transverse velocity at $9.0_{-2.2}^{+2.8}$ kpc.
The problem with that scenario is that J181503-375120 would spend only
a tiny fraction of its orbit near where it is observed today, and we
are therefore observing it at a special time.  There are two possible
interpretations of this observation.  The first is that both UCAC4 and
SPM4 have somehow overestimated the $\mu_{\alpha} \cos{\delta}$ proper
motion of J181503-375120.  This cannot easily be rejected.  Though the
UCAC4 and SPM4 proper motion measurements were produced independently,
they both used the same blue SPM plates for their first epoch astrometry.
In that case, the apparently large proper motion of J181503-375120
could be the result of an issue with the same blue SPM plate.  Moreover,
both UCAC4 and SPM4 may be subject to residual systematic uncertainties
at the level of 10 mas yr$^{-1}$.  The second interpretation is that
J181503-375120 is genuinely on a very eccentric orbit that takes it from
the bulge all the way to the edge of the Local Group.  Though we cannot
reject the latter hypothesis, we suspect that the former is a better
explanation.  Nevertheless, the proper motion of J181503-375120 merits
further attention.  If its parallax is measured and its proper motion
confirmed by Gaia, then it could be a hypervelocity star that has been
ejected from the Galactic center by a three-body interaction involving
the Milky Way's supermassive black hole.  In any case, J181503-375120
is currently located near the center of the Galaxy.

Since all three stars in our sample are old, one might wonder if the
orbits we observe today might be significantly different from their orbits
at higher redshift.  Even though we will argue that our stars formed at
$z \sim 10$, they were likely accreted by the Milky Way more recently.
\citet{tum10} found that even metal-poor stars that formed at $z \sim
10$ are not typically accreted by a Milky Way analog until $z \sim 3$.
\citet{wan11} showed that in the absence of a major merger, inside of
2 kpc Milky Way-analog dark matter halos have accreted more than 75\%
of their $z = 0$ mass by $z \sim 3$.  The Milky Way is not likely to
have had a major merger in that interval, as its disk is quite old and
its bulge appears to be a psuedobulge best explained by secular disk
instabilities \citep[e.g.,][]{aum09,sch09,kor04,how09}.  The fact that
the mass enclosed by the orbits of of our stars does not change much
since they likely entered the Milky Way's dark matter halo suggests
that their orbits should not have changed significantly.  The impact
of merger activity would be to cause the outward diffusion of stellar
orbits anyway, so in that situation the orbits of our stars would have
been even more tightly bound in the past.  This would not qualitatively
effect our interpretation of their abundances.

The inside-out formation of the Milky Way suggests that in the inner few
kpc of the Galaxy, about 10\% of stars with $\mathrm{[Fe/H]} \lesssim
-3.0$ formed at $z \gtrsim 15$ \citep{tum10}.  Another 20--40\% of stars
in the range $-4.0 \lesssim \mathrm{[Fe/H]} \lesssim -3.0$ formed at
$10 \lesssim z \lesssim 15$.  All three of our stars are currently in
the inner Galaxy, while the kinematics of two of the three place them
on tightly bound orbits.  The probability $P_{15}$ that at least one
of our stars formed at $z \gtrsim 15$ is 1 minus the probability that
none of them formed at $z \gtrsim 15$: $P_{15} = 1-0.9^3 \approx 0.3$.
Likewise, the probability $P_{10}$ that at least one of our stars formed
at $z \gtrsim 10$ is $P_{10} = 1-0.7^3 \approx 0.7$.  In other words,
there is 30\% chance that at least one of these three stars formed at
$z \gtrsim 15$ and a 70\% chance that at least one star formed at $10
\lesssim z \lesssim 15$.  If we apply the \citet{tum10} analysis only to
J155730-293922 and J183713-314109, then $P_{15} \approx 0.2$ and $P_{10}
\approx 0.5$.  Even though these stars are not the most metal-poor stars
known, the combination of their low metallicity and tightly-bound orbits
suggests that they may be among the most ancient stars with detailed
chemical abundance measurements.

In this scenario, our derived chemical abundances are indicative of the
chemical abundances of the progenitor galaxies of the Milky Way during
the epoch of the first galaxies.  Generally, we find that our abundance
ratios are near the mean of abundance distributions observed in halo
metal-poor giant stars \citep{cay04,yon13a,roe14}.  Si, Sc and Mn are
exceptions though, which we discuss below.  Based on four metal-poor bulge
stars with $-2.7 \lesssim \mathrm{[Fe/H]} \lesssim -2.5$ from the EMBLA
survey, \citet{how14} reached a similar conclusion: bulge metal-poor stars
have a similar abundance pattern to halo metal-poor stars.  They also
noted large scatter in [\ion{Mg}{1}/Fe] from $-0.07$ to $+0.62$ in just
four stars, with one star overabundant in [\ion{Ti}{2}/Fe] to the level
of $+0.84$. We find very little variance in [\ion{Mg}{1}/Fe], ranging
from $+0.46$ to $+0.57$.  None of our stars are overabundant in either
[\ion{Ti}{1}/Fe], [\ion{Ti}{2}/Fe], or any other $\alpha$-elements.
In fact, we find that [\ion{Ti}{1}/Fe] and [\ion{Ti}{2}/Fe] are about
$0.15$\,dex below the abundances of other $\alpha$-elements.

Our stars appear near the extremes of the silicon
abundance distribution observed in halo metal-poor giant stars
\citep[e.g.,][]{cay04,yon13a,roe14}.  This is likely due to their low
surface gravities and temperatures though.  \citet{bon09} found that
giants exhibited higher [Si/Fe] abundance ratios than dwarfs by about
0.2\,dex.  Similarly, cool stars usually appear to have high silicon
\citep{pre06,lai08,yon13a}. Given these two effects, the slightly higher
[Si/Fe] abundance ratios we find can most likely be attributed to a
combination of low surface gravity and cool temperatures. Indeed, when
we consider [Si/Fe] in giant stars ($\log{g} < 3$) in the \citet{roe14}
sample, our [Si/Fe] ratios lie near the mean for our temperature
range. That is to say although our stars show relatively high [Si/Fe]
ratios, the stars with high [Si/Fe] values in the comparison samples
also usually have cooler temperatures.  In short, [Si/Fe] appears to be
strongly correlated with temperature.  On the other hand, high silicon
is consistent with the Galactic chemical enrichment model predictions of
\citet{kob06}.  While we regard the former as the most likely explanation,
we cannot rule out the latter idea that the high silicon we observe is
representative of the $z \gtrsim 10$ interstellar medium.

The [\ion{Mn}{1}/Fe] abundance ratios we find are lower than what is
observed in metal-poor giants in the halo.  We use the same hyperfine
structure data for \ion{Mn}{1} as the referenced authors and derive
abundances from common lines.  \citet{cay04} and \citet{roe10,roe14}
have noted that the \ion{Mn}{1} resonance triplet at 403\,nm
yields systematically lower abundances than other neutral Mn lines.
For that reason, \citet{roe14}\footnote{\citet{cay04} made similar
adjustments.} empirically corrected their \ion{Mn}{1} triplet abundances
by about +0.3\,dex, which explains most of the discrepancy we observe.
\citet{yon13a} made no corrections, and we still find our stars in the
lower envelope of their [\ion{Mn}{1}/Fe] distribution.  The remaining
difference in [\ion{Mn}{1}/Fe] is probably  attributable to our stars
being at the tip of the giant branch.  In halo metal-poor giant stars,
many authors have noted a positive trend in the $T_{\mathrm{eff}}-{\rm
[Mn/Fe]}$ plane.  In other words, lower [Mn/Fe] abundances are found in
cooler giants \citep{pre06,yon13a,roe14}.

\begin{figure}
%\epsscale{1.0} % aastex
\plotone{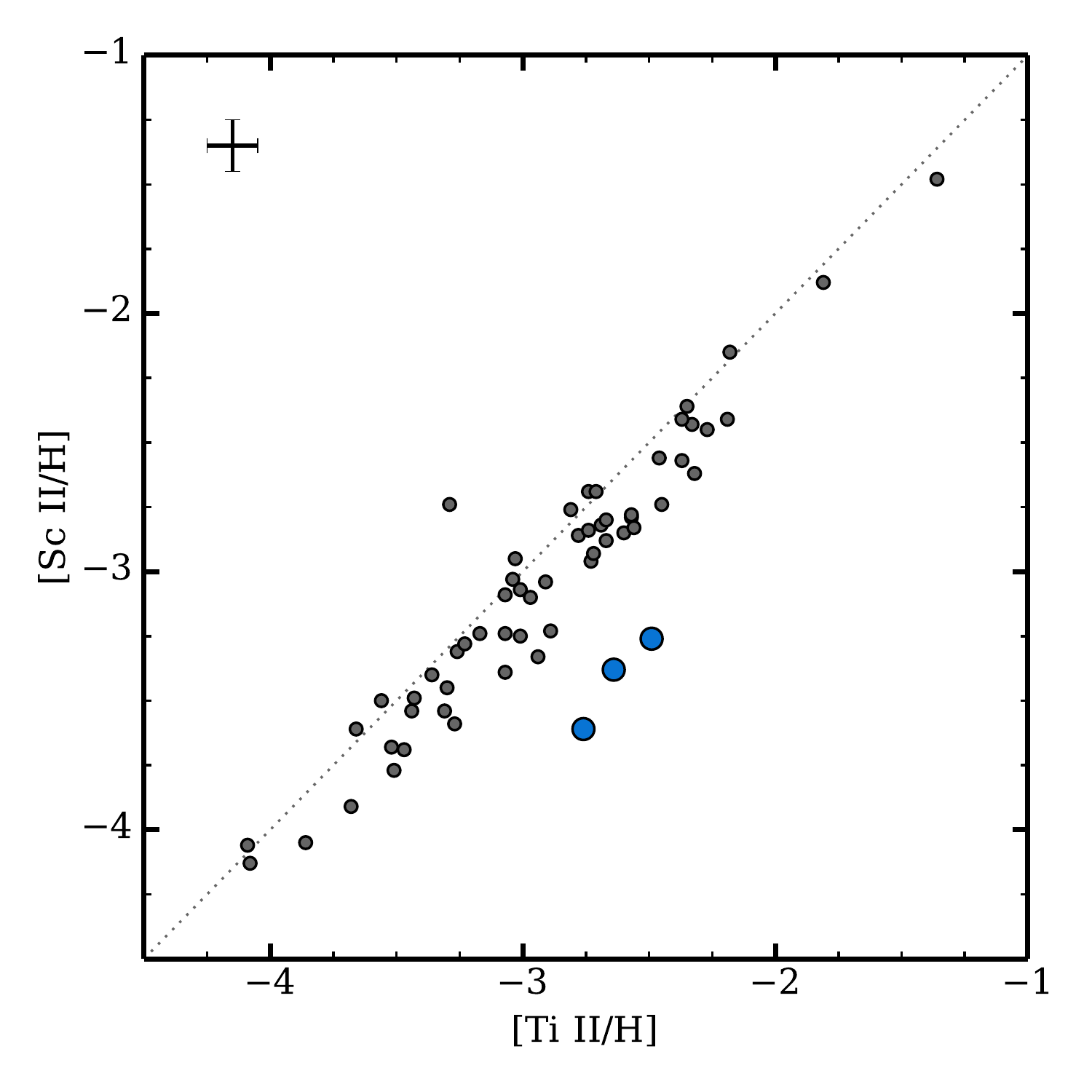}
\caption{Abundance ratios [\ion{Ti}{2}/H] versus [\ion{Sc}{2}/H].
We plot our three bulge metal-poor stars in blue and the \citet{yon13a}
comparison sample in dark gray.  Typical uncertainties are given.
Our stars in the bulge significantly deviate from this relation observed
in the halo.\label{fig06}}
\end{figure}

All three stars have low scandium abundances, with [\ion{Sc}{2}/Fe]
$\lesssim -0.5$.  \citet{yon13a} found a tight abundance relation between
[\ion{Ti}{2}/H] and [\ion{Sc}{2}/H] in halo stars, which is suggestive
of a common nucleosynthetic environment.  Figure~\ref{fig06} shows that
our stars deviate significantly from this relation. Unlike \ion{Si}{1}
or \ion{Mn}{1}, our low [\ion{Sc}{2}/Fe] abundance ratios cannot be
easily explained by correlations with $T_{\mathrm{eff}}$. \citet{yon13a}
found a slight slope ($m = 0.05 \pm 0.06$) in the relationship between
$T_{\mathrm{eff}}$ and [\ion{Sc}{2}/Fe], such that cooler stars
have lower [\ion{Sc}{2}/Fe] abundance ratios.  The typical range of
[\ion{Sc}{2}/Fe] they measure for cool stars is $-0.10$ to $+0.50$ though.
Our measurements are substantially below this range, with [\ion{Sc}{2}/Fe]
= $-0.59$ to $-0.54$.

Scandium probably remains the most discrepant element between Galactic
chemical evolution models and observations of metal-poor stars, as models
typically under-predict Sc abundances by a factor of ten.  For example,
\citet{kob06} predict constant ${[{\rm Sc/Fe}] \sim -1}$ for metal-poor
stars, roughly an order of magnitude lower than the observed values of
[Sc/Fe] $\sim 0$.  The abundance ratios we find in the inner few kpc of
the Galaxy bring our stars far closer to these predictions.  However,
advances in modeling are required for both abundance measurements
(e.g., non-LTE treatment, $\langle$3D$\rangle$ photospheres) and
Galactic chemical evolution models.  Departures from local thermodynamic
equilibrium or 3D effects will alter the inferred Sc abundances, while
increasing the $\alpha$-rich freeze-out or delaying neutrino processes
during explosive nucleosynthesis may be necessary to increase Sc yields
in chemical evolution models \citep[e.g.,][]{fro06,kob06}.

We searched the SAGA database\footnote{Described in
\citet{sud08,sud11} and \citet{yam13} and available at
\url{http://saga.sci.hokudai.ac.jp/wiki/doku.php}.} and the compilation
of \citet{fre10a} for other Galactic giant stars with [\ion{Sc}{2}/Fe]
$\lesssim -0.5$.  That search returned three objects: BS 16929-005,
HE 0533-5340, and HE 1207-3108.  While BS 16929-005 was reported by
\citet{hon04} to have [\ion{Sc}{2}/Fe] $= -0.53$, that measurement
did not take into account the hyperfine structure that is known to be
important for scandium abundance measurements \citep[e.g.,][]{pro00}.
In comparison, \citet{lai08} accounted for hyperfine structure in BS
16929-005 and found [\ion{Sc}{2}/Fe] $= -0.03$.  We regard the latter
measurement as more reliable.  \citet{coh13} found [\ion{Sc}{2}/Fe] $=
-0.56$ for HE 0533-5340 and \citet{yon13a} found [\ion{Sc}{2}/Fe] $=
-0.55$ for HE 1207-3108.  However, both HE 0533-5340 and HE 1207-3108 are
among the rare class of ``iron-rich" metal-poor stars in which most [X/Fe]
abundances are sub-solar.  This is in contrast to typical metal-poor
stars, which are usually enhanced in at least the $\alpha$ elements.
The combination of low [\ion{Sc}{2}/Fe] and $\alpha$ enhancement that we
see in our three metal-poor giants in the bulge is unprecedented in any
of the 381 metal-poor giant stars in the SAGA database with scandium
abundance measurements.  These three stars are therefore unlike any
other known star in the Galaxy.

We also searched \citet{fre10a} for metal-poor stars in dwarf galaxies
with [\ion{Sc}{2}/Fe] $\lesssim -0.5$.  We found two examples, one
from \citet{fre10b} in Coma Berenices (SDSS J122657+235611/ComBer-S3)
and one from \citet{she03} in Carina (Car 3).  Car 3 is an ``iron-rich"
metal-poor star, so we do not consider it further.  That leaves SDSS
J122657+235611/ComBer-S3 with [\ion{Sc}{2}/Fe] $= -0.57$ as the only
giant star known with a similar abundance pattern to our three metal-poor
giants in the bulge.  Coma Berenices is an ultra-faint dwarf spheroidal
(dSph) galaxy with a $V$-band absolute magnitude of only $M_{V} = -3.4$
\citep{bel07,dej08}.  It is also one of the most ancient galaxies known.
Indeed, \citet{bro14} found a mean age of $13.9 \pm 0.3$ Gyr for Coma
Berenices based on \textit{Hubble Space Telescope} Advanced Camera
for Surveys photometry of its resolved stellar population.  That made
it the oldest galaxy in their sample.  The apparent chemical abundance
similarity between the ancient dSph Coma Berenices and our three stars in
the bulge supports both the conclusion that our three stars are among the
most ancient stars in our Galaxy and the idea that low [\ion{Sc}{2}/Fe]
may be a chemical indicator of ancient stellar populations.

Our detailed chemical abundance analysis has assumed that transition
levels are in a state of LTE. It is well known that this assumption
breaks down in the upper levels of stellar photospheres, where
departures from LTE can significantly alter the inferred elemental
abundance. The direction and magnitude of these abundance changes
are dependent on stellar parameters, atomic number, ionization
level, absorption depth (i.e., the strength of the transition),
among other factors.  Many authors have investigated the effects
of abundance deviations due to LTE departures in well-studied
metal-poor giant stars that are comparable to our program stars,
like HD\,122563 \citep[e.g.,][]{gra99,asp03,mas08,and10,han13}.
For metal-poor giant stars like those analyzed here, the abundance
changes due to departures from LTE will be the largest for \ion{K}{1},
\ion{Co}{1}, and \ion{Mn}{1}. The change in \ion{K}{1} is significantly
negative\footnote{Deviations are described following standard
nomenclature: $\Delta{\rm NLTE} = \log_{\rm \epsilon}({\rm X})_{\rm NLTE}
- \log_{\rm \epsilon}({\rm X})_{\rm LTE}$.  A `positive correction'
refers to a higher abundance after accounting for departures from LTE.}:
$\Delta\log_{\rm \epsilon}$\ion{K}{1} $\approx -0.15$, such that in
Figure \ref{fig03} we have shown uncorrected (i.e., LTE) \ion{K}{1}
abundances from \citet{roe14} for a fair comparison. \ion{Co}{1} is
expected to show the largest absolute change, with positive deviations
up to about $+0.65$\,dex. Similarly we can expect our \ion{Mn}{1}
abundances to increase by about $+0.4$\,dex with the proper inclusion of
LTE departure coefficients. However, these \ion{Mn}{1} corrections would
be of the same approximate order and direction for the halo comparison
samples. Therefore we assert that the \ion{Mn}{1} abundance ratios
we find in metal-poor stars in the bulge would persist in the lower
tail of [\ion{Mn}{1}/Fe] abundance distribution observed in comparable
halo stars.  All other species examined here have expected abundance
deviations less than $0.2$\,dex, with the average magnitude being about
$0.1$\,dex \citep{ber14}.  We note that systematic abundance differences
can also be expected due to surface granulation and convection, complex
features which cannot be accounted for in our 1D models.

Our observations indirectly suggest that the progenitor galaxies of
the Milky Way had reached $\mathrm{[Fe/H]} \sim -3.0$ with an abundance
pattern comparable to metal-poor halo stars by $z \sim 10$.  The chemical
state of high-redshift galaxies can be measured directly by observations
of metal-poor damped Ly$\alpha$ systems (DLAs) in absorption in the
spectra of background quasars.  Many authors\footnote{See for example
\citet{mol00}, \citet{des01}, \citet{pro02}, \citet{des03}, \citet{ome06},
\citet{peti08}, \citet{pett08}, \citet{ell10}, \citet{pen10},
\citet{sri10}, and \citet{coo11a,coo11b}.} have measured the column
densities and relative abundances of H, C, N, O, Al, Si, and Fe to $z
\sim 4$.   At higher redshift, C, O, Mg, Si, and Fe have been measured
in DLAs at $z \sim 6$ \citep{bec12}.  At $z \approx 7$, the abundances of
one system has been bounded to be less than 1/1,000 solar \citep{sim12}.
Where [C/Fe], [O/Fe], and [Si/Fe] have been measured in high-redshift
DLAs, it has been found that the average abundances are in good agreement
with those observed in metal-poor stars: $\mathrm{[C/Fe]} \approx 0.15
\pm 0.03$, $\mathrm{[O/Fe]} \approx 0.40 \pm 0.01$, and $\mathrm{[Si/Fe]}
\approx 0.37 \pm 0.01$.  Our stars in the bulge are likely ancient and
are well matched by the observed abundances in DLAs.  Only 500 Myr passes
between $z \sim 10$ and $z \sim 6$ \citep[e.g.,][]{wri06}, so it seems
plausible that the $z \sim 10$ abundances as observed in our ancient
stars (after correcting for $\log{g}$ and $T_{\mathrm{eff}}$ effects)
are comparable to those directly observed at $z \sim 6$.

\section{Conclusions}\label{sec:conclusions}

We have measured the detailed chemical abundances of the three metal-poor
stars with $\mathrm{[Fe/H]} \lesssim -2.7$ in the bulge that we discovered
in \citet{sch14}.  Two of these three stars are the most metal-poor
stars in the bulge in the literature, while the third is comparable to
the most metal-poor star identified in \citet{how14}.  We have carefully
estimated the Galactocentric distances and orbits of all three stars.
While we find that all three have $d_{\mathrm{gc}} \lesssim 4$\,kpc, only
J155730-293922 and J183713-314109 can be securely placed on tightly-bound
orbits.  J181503-375120 may be a halo star on a very eccentric orbit that
is only passing through the bulge.  While UCAC4 and SPM4 proper motion
measurements favor a very eccentric orbit, the orbit is so extreme that
it may be more likely that there is an issue with the SPM blue plate
that provides the first epoch astrometry for both catalogs. When combined
with their metal-poor nature, their proximity to the center of the Galaxy
and their tightly-bound orbits indicate that these stars  may be some of
the most ancient objects yet identified.  We use the theoretical models
of \citet{tum10} to estimate that there is a 30\% chance that at least
one of these stars formed at $z \gtrsim 15$ and a 70\% chance that at
least one formed at $10 \lesssim z \lesssim 15$.  We therefore argue
that the chemical abundances we observe in these metal-poor stars is
representative of the chemical state of the interstellar medium in the
progenitor galaxies of the Milky Way at $z \sim 10$.

Compared to observations of metal-poor giant stars of similar effective
temperatures found in the Galactic halo, we find similar [X/Fe]
abundance ratios for most elements.  However, we observe [\ion{Sc}{2}/Fe]
abundance ratios lower than reported in the halo by about $0.5$\,dex.
Scandium remains the element with the largest discrepancy between what
is observed in halo metal-poor stars and what is predicted from models of
Galactic chemical evolution.  Interestingly, when compared to the values
observed in halo metal-poor stars, our [\ion{Sc}{2}/Fe] abundances are
closer to predictions for the chemical abundances of the first galaxies
\citep[e.g.,][]{kob06}.  For these reasons, the progenitor halos of the
Milky Way likely reached $\mathrm{[Fe/H]} \sim -3.0$ by $z \sim 10$.
Their chemical abundances were probably very similar to those observed
in halo metal-poor stars with the possible exception of Sc, which we
observe to be low in these ancient stars in the bulge.

\acknowledgments
We thank Judith Cohen, Anna Frebel, Gerry Gilmore, Paul Schechter, and
Josh Winn.  We are especially grateful to the anonymous referee for
suggestions that improved this paper.  This research has made use
of NASA's Astrophysics Data System Bibliographic Services and both
the SIMBAD database and VizieR catalog access tool, CDS, Strasbourg,
France.  The original description of the VizieR service was published by
\citet{och00}.  This research made use of Astropy, a community-developed
core Python package for Astronomy \citep{astropy}.  This publication
makes use of data products from the Wide-field Infrared Survey Explorer,
which is a joint project of the University of California, Los Angeles, and
the Jet Propulsion Laboratory/California Institute of Technology, funded
by the National Aeronautics and Space Administration.  This publication
makes use of data products from the Two Micron All Sky Survey, which
is a joint project of the University of Massachusetts and the Infrared
Processing and Analysis Center/California Institute of Technology, funded
by the National Aeronautics and Space Administration and the National
Science Foundation.  This publication was partially based on observations
obtained at the Gemini Observatory, which is operated by the Association
of Universities for Research in Astronomy, Inc., under a cooperative
agreement with the NSF on behalf of the Gemini partnership: the National
Science Foundation (United States), the National Research Council
(Canada), CONICYT (Chile), the Australian Research Council (Australia),
Minist\'{e}rio da Ci\^{e}ncia, Tecnologia e Inova\c{c}\~{a}o (Brazil)
and Ministerio de Ciencia, Tecnolog\'{i}a e Innovaci\'{o}n Productiva
(Argentina).  This research has made use of the NASA/IPAC Infrared
Science Archive, which is operated by the Jet Propulsion Laboratory,
California Institute of Technology, under contract with the National
Aeronautics and Space Administration.  This research was made possible
through the use of the AAVSO Photometric All-Sky Survey (APASS), funded
by the Robert Martin Ayers Sciences Fund. A.~R.~C. acknowledges support
through European Research Council grant 320360: The Gaia-ESO Milky Way
Survey.  Support for this work was provided by the MIT Kavli Institute for
Astrophysics and Space Research through a Kavli Postdoctoral Fellowship.

{\it Facilities:} \facility{CTIO:2MASS}, \facility{FLWO:2MASS}, 
\facility{Gemini:South (GMOS-S spectrograph)}, \facility{Magellan:Clay
(MIKE spectrograph)}, \facility{WISE}

\clearpage

\clearpage

\clearpage

\clearpage

\clearpage

\clearpage

\clearpage
\tabletypesize{\small}
\begin{deluxetable}{lccrrccccccc}
%\rotate %aastex
\tablecaption{Bulge Extremely Metal-poor Star Positions and Photometry\label{tbl-1}}
\tablewidth{0pt}
\tablehead{\colhead{Object} & \colhead{RA} & \colhead{DEC} & \colhead{$l$} & \colhead{$b$} & \colhead{$V$} &
\colhead{$B-V$} & \colhead{$J$} & \colhead{$H$} & \colhead{$K_{s}$} & \colhead{$W1$} & \colhead{$W2$} \\
\colhead{(2MASS)} & \omit & \omit & \colhead{(deg)} & \colhead{(deg)} & \colhead{(mag)} &
\colhead{(mag)} & \colhead{(mag)} & \colhead{(mag)} & \colhead{(mag)} & \colhead{(mag)} & \colhead{(mag)}} 
\startdata
J155730.10$-$293922.7 & 15:57:30.1 & -29:39:23 & 344 &  18 & 13.15 & 0.97  & 11.13 & 10.61 & 10.50 & 10.39 & 10.40 \\
J181503.64$-$375120.7 & 18:15:03.6 & -37:51:20 & 355 & -10 & 12.87 & 1.01  & 10.80 & 10.29 & 10.19 & 10.12 & 10.13 \\
J183713.28$-$314109.3 & 18:37:13.2 & -31:41:09 &   3 & -11 & 12.47 & 0.82  & 10.65 & 10.12 & 10.04 &   9.90 &   9.90
\enddata
%\tablecomments{}
\end{deluxetable}

\clearpage
\tabletypesize{\small}
\begin{deluxetable}{lrrrrr}
%\rotate %aastex
\tablecaption{Bulge Extremely Metal-poor Star Kinematic Observables\label{tbl-2}}
\tablewidth{0pt}
\tablehead{\colhead{Object} & \colhead{$v_{\rm hel}$} &
\colhead{$\mu_{\alpha} \cos{\delta}$\tablenotemark{a}} & \colhead{$\mu_{\delta}$\tablenotemark{a}} &
\colhead{$\mu_{\alpha} \cos{\delta}$\tablenotemark{b}} & \colhead{$\mu_{\delta}$\tablenotemark{b}} \\
\colhead{(2MASS)} & \colhead{(km s$^{-1}$)} & \colhead{(mas yr$^{-1}$)} &
\colhead{(mas yr$^{-1}$)} & \colhead{(mas yr$^{-1}$)} &
\colhead{(mas yr$^{-1}$)}}
\startdata
J155730.10$-$293922.7 & $ 134.0\,\pm\,1.0$ &   $1.5\,\pm\,1.4$ &  $5.9\,\pm\,1.4$ & $-5.68\,\pm\,2.24$ & $5.26\,\pm\,2.10$ \\
J181503.64$-$375120.7 & $ -77.8\,\pm\,1.0$ & $-19.6\,\pm\,2.1$ & $-2.9\,\pm\,2.2$ &$-20.11\,\pm\,3.37$ & $5.47\,\pm\,3.21$ \\
J183713.28$-$314109.3 & $-199.0\,\pm\,1.0$ & $-11.4\,\pm\,1.7$ & $-2.5\,\pm\,1.7$ & $-8.31\,\pm\,3.18$ & $-7.04\,\pm\,3.21$
\enddata
%\tablecomments{}
\tablenotetext{a}{UCAC4 proper motions from \citet{zac13}}
\tablenotetext{b}{SPM4 proper motions from \citet{gir11}}
\end{deluxetable}

\clearpage
\tabletypesize{\small}
\begin{deluxetable}{lcccc}
\tablecaption{Stellar Parameters of Bulge Extremely Metal-poor Stars.\label{tbl-3}}
\tablewidth{0pt}
\tablehead{ 
\colhead{Object} & \colhead{$T_{\rm eff}$} &
\colhead{$\log{g}$} & \colhead{[Fe/H]} & \colhead{$\xi$} \\
\colhead{(2MASS)} & \colhead{(K)} & \colhead{(cm s$^{-2}$)} & \colhead{} & \colhead{(km s$^{-1}$)}}
\startdata
J155730.10$-$293922.7 & 4720  & 1.12  & $-3.02$ & 2.88 \\
J181503.64$-$375120.7 & 4728  & 1.09  & $-2.84$ & 3.00 \\
J183713.28$-$314109.3 & 4797  & 0.99  & $-2.70$ & 2.67
\enddata
\tablecomments{We estimate the uncertainties in $T_{\mathrm{eff}}$,
$\log{g}$, [Fe/H], and $\xi$ to be 100\,K, 0.2\,dex, 0.1\,dex, and
0.1\,km s$^{-1}$.}
\end{deluxetable}

\clearpage
\tabletypesize{\small}
\begin{deluxetable}{lcccrccrc}
%\rotate %aastex
\tablecaption{Bulge Extremely Metal-poor Star Derived Properties\label{tbl-4}}
\tablewidth{0pt}
\tablehead{\colhead{Object} & \colhead{$d_{\odot}$} & \colhead{$d_{\mathrm{gc}}$} &
\colhead{$r_{\mathrm{peri}}$\tablenotemark{a}} &
\colhead{$r_{\mathrm{ap}}$\tablenotemark{a}} & \colhead{$e$\tablenotemark{a}} &
\colhead{$r_{\mathrm{peri}}$\tablenotemark{b}} &
\colhead{$r_{\mathrm{ap}}$\tablenotemark{b}} & \colhead{$e$\tablenotemark{b}} \\
\colhead{(2MASS)} & \colhead{(kpc)} & \colhead{(kpc)} &
\colhead{(kpc)} & \colhead{(kpc)} & \omit &
\colhead{(kpc)} & \colhead{(kpc)} & \omit}
\startdata
J155730.10$-$293922.7 & $9.8_{-2.3}^{+3.0}$ & $3.4_{-1.6}^{+2.2}$ & $2.9_{-2.0}^{+3.3}$ & $16_{-6.8}^{+17}$ & $0.747_{-0.243}^{+0.137}$ & $1.7_{-1.0}^{+3.1}$  & $7.3_{-3.2}^{+22}$ & $0.682_{-0.269}^{+0.194}$ \\
J181503.64$-$375120.7 & $9.0_{-2.2}^{+2.8}$ & $3.2_{-1.4}^{+1.5}$ & $2.1_{-0.50}^{+1.9}$  & $4900_{-3800}^{+3600}$ & $0.999_{-0.003}^{+0.000}$ & $2.1_{-0.57}^{+1.9}$  & $5400_{-4200}^{+4000}$ & $0.999_{-0.002}^{+0.000}$ \\
J183713.28$-$314109.3 & $9.6_{-2.3}^{+3.1}$ & $3.3_{-1.5}^{+2.0}$ & $1.8_{-1.1}^{+2.7}$ & $88_{-72}^{+3400}$ & $0.979_{-0.138}^{+0.018}$ & $1.4_{-0.87}^{+2.6}$ & $17_{-12}^{+210}$ & $0.888_{-0.201}^{+0.093}$
\enddata
%\tablecomments{}
\tablenotetext{a}{Using UCAC4 proper motions.}
\tablenotetext{b}{Using SPM4 proper motions.}
\end{deluxetable}

\clearpage
\LongTables % emulateapj
\tabletypesize{\small}
\begin{deluxetable}{cllrrrcrrcrr}
\tablecaption{Measured Equivalent Widths and Abundances of Bulge Extremely Metal-poor Stars.\label{tbl-5}}
\tablewidth{0pt}
\tablehead{
& & & & \multicolumn{2}{c}{J155730$-$293922} && \multicolumn{2}{c}{J181503$-$375120} && \multicolumn{2}{c}{J183713$-$314109} \\
\cline{5-6} \cline{8-9} \cline{11-12} 
\colhead{Wavelength} & \colhead{Species} & \colhead{$\chi$} & \colhead{$\log{gf}$} & \colhead{EW} & \colhead{$\log_{\epsilon}{\rm X}$} && \colhead{EW} & \colhead{$\log_{\epsilon}{\rm X}$} && \colhead{EW} & \colhead{$\log_{\epsilon}{\rm X}$} \\
\colhead{(\AA)} & & \colhead{(eV)} & & \colhead{(m\AA)} &&& \colhead{(m\AA)} &&& \colhead{(m\AA)}}
\startdata
% Since Li is only determined from one feature, and that feature is noted in text, and the abundances are noted in another table, it is probably superfluous to list it here. Commented out temporarily:
%$\sim$6707\tablenotemark{a} & Li I & 0.00 & \nodata & \nodata & 0.25 && \nodata & 0.40 && \nodata & 0.10 \\
5889.95 & \ion{Na}{1}  & 0.00 &   +0.11 & 157.8 & 3.40 && 170.4 & 3.52 && 182.7 & 3.96 \\
5895.92 & \ion{Na}{1}  & 0.00 & $-$0.19 & 139.0 & 3.39 && 144.0 & 3.41 && 158.2 & 3.87 \\
3829.36 & \ion{Mg}{1}  & 2.71 & $-$0.21 & 156.0 & 5.21 && 169.6 & 5.32 && \nodata & \nodata \\
3832.30 & \ion{Mg}{1}  & 2.71 &   +0.27 & 191.9 & 5.13 && \nodata & \nodata && 197.3 & 5.30 \\
3838.29 & \ion{Mg}{1}  & 2.72 &   +0.49 & 220.0 & 5.13 && \nodata & \nodata && 229.1 & 5.33 \\
3986.75 & \ion{Mg}{1}  & 4.35 & $-$1.03 & 22.9 & 5.22 && 40.8 & 5.59 && 32.7 & 5.46 \\
4057.51 & \ion{Mg}{1}  & 4.35 & $-$0.89 & \nodata & \nodata && 34.5 & 5.31 && 39.1 & 5.43 \\
4167.27 & \ion{Mg}{1}  & 4.35 & $-$0.71 & 41.6 & 5.24 && 55.8 & 5.47 && 45.2 & 5.34 \\
\enddata
\tablenotetext{a}{Abundance determined by synthesis with relevant
hyperfine and/or isotopic splitting data included where appropriate. See
text for details.}
\tablecomments{The colon symbol for \ion{Cu}{1} in J181503.64$-$375120.7
indicates high uncertainty in this measurement. See text for details.}
\end{deluxetable}

\clearpage
%\LongTables % emulateapj
\tabletypesize{\small}
\begin{deluxetable}{lcrrrr}
\tablecaption{Chemical Abundances of Bulge Extremely Metal-poor Stars.\label{tbl-6}}
\tablewidth{0pt}
\tablehead{\colhead{Species} & \colhead{$N$} & \colhead{$\log_{\rm \epsilon}\left({\rm X}\right)$} & [X/H] & [X/Fe] & $\sigma_{\rm lines}$}
\startdata
& \\
\multicolumn{6}{c}{2MASS J155730.10$-$293922.7} \\
\hline
\ion{Li}{1} &  1 &    +0.25 &  $-$0.80 &   +2.22 & \nodata \\
C (CH)      &  1 &    +5.41 &  $-$3.02 &   +0.00 & \nodata \\
\ion{Na}{1} &  2 &    +3.40 &  $-$2.85 &   +0.18 & $<$0.01 \\
\ion{Mg}{1} & 10 &    +5.12 &  $-$2.48 &   +0.54 &    0.08 \\
\ion{Al}{1} &  1 &    +2.73 &  $-$3.72 & $-$0.70 & \nodata \\           
\ion{Si}{1} &  1 &    +5.29 &  $-$2.22 &   +0.81 & \nodata \\
\enddata
\tablecomments{The colon symbol for \ion{Cu}{1} in J181503.64$-$375120.7 indicates high uncertainty in this measurement. See text for details.}
\end{deluxetable}

\clearpage
\LongTables % emulateapj
\tabletypesize{\scriptsize}
\begin{deluxetable}{lcccccccc}
%\rotate %aastex
\tablecaption{Systematic, Random, and Total Chemical Abundance Uncertainties of Bulge Extremely Metal-poor Stars. \label{tbl-7}}
\tablewidth{0pt}
\tablehead{%& & & & & \multicolumn{2}{c}{Total Uncertainty} \\ \cline{6-7}
& \multicolumn{4}{c}{Systematic Uncertainty} && \colhead{Random Uncertainty} && \colhead{Total Uncertainty} \\ \cline{2-5} \cline{7-7} \cline{9-9}
\colhead{Species ([X/H])} & \colhead{$T_{\rm eff} +100\,{\rm K}$} & \colhead{$\log{g} +0.20\,{\rm dex}$} & \colhead{$\xi +0.10$\,km s$^{-1}$} & \colhead{${\rm [M/H]} +0.10\,{\rm dex}$} && \colhead{$\max(0.10, \sigma_{\rm lines})/\sqrt{N}$} && \colhead{$\sigma_{\rm total}$} \\
& \colhead{$\Delta$abund. (dex)} & \colhead{$\Delta$abund. (dex)} & \colhead{$\Delta$abund. (dex)} & \colhead{$\Delta$abund. (dex)} && \colhead{(dex)} && \colhead{(dex)}}
\startdata
& \\
\multicolumn{9}{c}{2MASS J155730.10$-$293922.7} \\
\hline
 Na I &             +0.15 &            $-$0.02 &          $-$0.04 &             $-$0.01 &&  0.07 && 0.17 \\
 Mg I &             +0.08 &            $-$0.08 &          $-$0.02 &             $-$0.01 &&  0.03 && 0.12 \\
 Al I &             +0.04 &            $-$0.11 &          $-$0.05 &             $-$0.02 &&  0.10 && 0.16 \\
 Si I &             +0.08 &            $-$0.09 &          $-$0.03 &             $-$0.01 &&  0.07 && 0.14 \\
  K I &             +0.09 &            $-$0.01 &          $-$0.01 &               +0.00 &&  0.08 && 0.12 \\
 Ca I &             +0.06 &            $-$0.03 &          $-$0.01 &             $-$0.00 &&  0.02 && 0.07 \\
\enddata
\tablecomments{Total uncertainty ($\sigma_{\rm total}$) refers to the
quadrature sum of random uncertainties and systematic uncertainties in
stellar parameters ($T_{\rm eff}$, $\log{g}$, ${\rm [M/H]}$, $\xi$), where
the stellar parameter uncertainties are assumed to be uncorrelated. The
individual line uncertainties ($\sigma_{\rm lines}$) are assumed to be
at least 0.10\,dex for each species, such that the random uncertainties
are given by $\max(0.10,\sigma_{\rm lines})/\sqrt{N}$ where $N$ is the
number of transitions.}
\end{deluxetable}

\end{document}